\documentclass[12pt]{article}

\usepackage{amsmath,amsfonts,amssymb}

\newcommand{\half}{\frac{1}{2}}

\newcommand{\imi}{\mathrm{i}}
\newcommand{\dex}{\mathbin{\mathrm{d}}}

\newcommand{\ph}{\varphi}
\newcommand{\xx}{\tilde x}
\newcommand{\yy}{\tilde y}

\newcommand{\mykappa}{{\varkappa}} 

\newcommand{\pp}{{\epsilon}} 

\newcommand{\Ep}{{E_\#}} %
\newcommand{\dEp}{{E_\#{}'}} %

\newcommand{\Eq}{{%
Eq.~}}
\newcommand{\Eqs}{{%
Eqs.~}}

\newcommand{\JJ}{{Josephson junction}}
\newcommand{\DCHE}{{double confluent Heun equation}}

\begin{document}

\begin{center}{\large
General solution\\
of overdamped Josephson junction equation\\
in the case of phase-lock
}\\[2em]

{S.I.\ Tertychniy}

{VNIIFTRI, Mendeleevo, Moscow Region, 141570, Russia
}
\end{center}

\begin{abstract}
\noindent
 The first order nonlinear ODE 
 $\dot \ph(t) + \sin\ph(t)=B+A\cos\omega t,$ 
 ($ A,B,\omega$ are real constants)
 which is commonly used as a simple model 
 of an overdamped \JJ{} in superconductors
 is investigated.
Its general solution 
is obtained 
in the case of 
the choice of parameters associated with
one of three possible
kinds of asymptotic behavior of solutions
known as phase-lock
where all but one solutions 
converge to a common `essentially periodic' attractor.
The general solution is represented in explicit form 
in terms of the Floquet solution of a particular instance 
of the \DCHE{} (DCHE).
In turn, the solution of DCHE
is represented through the Laurent series which defines
an analytic 
function on the Riemann sphere with punctured poles.
The Laurent series coefficients are given in explicit form
in terms of
infinite  products
of $2\times2$ matrices with a single zero element.
The closed form of  the 
phase-lock condition
is obtained which is 
represented
as the condition of existence of a real root of 
the transcendental function.
The phase-lock criterion 
is conjectured
whose plausibility is confirmed in numerical tests.
\end{abstract}

\section{Introduction}

The nonlinear first order ODE 
 \begin{equation}
 \dot \ph(t) + \sin\ph(t)=q(t),
                                  \label{eq::1}
\end{equation}
is commonly used in applied physics 
as the simple mathematical model describing the
electric properties of 
\JJ{}s (JJ's) in
superconductors \cite{R1,R2}.
Here the
\textsc{rhs} function 
 $q(t)$
assumed to be known
specifies the external impact to JJ
representing
the appropriately normalized 
\textit{bias current} (or simply \textit{bias})
supplied by an external current source.
The unknown real valued  function 
$\ph(t)$ called {\it the phase\/}
describes the macroscopic quantum state of JJ. 
In particular, it is connected with the instantaneous
voltage $V$ applied across JJ 
in accord with 
the equation $V=(\hbar/2e)\hbox{$\dex\ph/\mathrm{d}\tau$}$, 
where $\hbar$ is the
Plank constant, $e$ is the electron charge,  $\tau$ is the
(dimensional) current time.
The dimensionless variable $t$ entering \Eq(\ref{eq::1}) 
is defined as $t=\omega_c\tau$, where
$\omega_c$ is a constant parameter 
depending on the junction properties and
named JJ characteristic frequency. 
See Ref \cite{R5} for more details of JJ physics.

\Eq(\ref{eq::1}) arises as the limiting case 
of the second order ODE utilized in 
more general Resistively Shunted Junction (RSJ) 
model  \cite{R3,R4}. The reduction is legitimate
if 
the role of the
junction capacitance proves negligible.
In practice, if 
JJ can be 
described by \Eq(\ref{eq::1}) it is named {\it overdamped}.
Summarizing the aforementioned relationships, we shall name \Eq(\ref{eq::1}),
for brevity,
 {\em overdamped \JJ{} equation} (OJJE).

Under concordant
conditions,
the theoretical modeling
applying
OJJE 
is in  excellent agreement with
experiments.
It is also 
worth noting 
that nowadays 
electronic devices based on 
the Josephson effect 
play the important role 
in  various branches of  measurement technology.
In particular,  JJ arrays
serve
the heart element of the modern DC voltage standards \cite{R6}.
The development of JJ-based synthesizers of AC voltage waveforms 
is currently in progress \cite{R7,R8,R9}.
These 
and other successful applications
stimulate 
the growing interest to 
the theoretical study and the modeling of JJ properties including
investigation of capabilities of RSJ model 
and its limiting cases
and the predictions they lead to.

To be more specific, 
the case most 
important
from viewpoint of applications 
and simultaneously 
distinguished by
the wealthiness 
of the underlaid mathematics
is definitely the one of the bias 
function representing harmonic oscillations.
Without loss of generality, 
it 
can be get in the following form
 \begin{equation}\label{eq::2} 
 q(t)=B+A\cos\omega t,
\end{equation}
where
$A,B,\omega>0$ are some real constant parameters.
Hereinafter, the abbreviation OJJE introduced above
will refer to the  couple (\ref{eq::1}),(\ref{eq::2}).

In spite of apparent simplicity of OJJE, 
few facts of its specific
analytic theory 
had been available until recently.
Some preliminary 
results concerning
the problem 
of derivation of analytic solutions
of OJJE 
in general setting
had been obtained in Ref.~\cite{R11a}.
The 
approach put forward
therein  
is elaborated in the present work.
The focus is made 
on the case of 
manifestation of the 
{\em phase-lock\/} property 
by \Eq(\ref{eq::1}) 
which is one of its most important features from viewpoint of applications. 
The phase-lock property is formalized as follows:
in the case of phase-lock
any solution $\varphi=\varphi(t)$ of 
\Eq(\ref{eq::1}) 
either yields a 
periodic exponent $e^{\imi\varphi}$,
exactly 
two such distinguished solutions existing, 
or, as the time parameter grows,
$e^{\imi\varphi}$ exponentially converges
to the similar exponent 
 for the one (common for all
$\varphi$'s)
 of the periodic functions  just noted (another one is the repeller).
(Here and in what follows 
we shall not distinguish 
phase functions which 
differ by a constant equal to $2\pi$ times an integer.)
The corresponding period coincides
with one of $q(t)$, i.e.,
in the case (\ref{eq::2}),
$2\pi\omega^{-1}$.
This behavior is stable with respect to  weak parameter perturbations,
i.e.\ the subset of parameter values leading to phase-lock
 is open.
It is worth noting for completeness that in the opposite
(no phase-lock) case
no {\it stable\/} periodicity 
in the behavior of $e^{\imi\varphi}$ 
is observed. There is also
a third, intermediate type of the phase behavior,
where the attractor and repeller are, in a sense, merged.
It is
realized on the lower-dimensional  subset of the space of
parameter values~\cite{R11}.

In the present work,
the complete analytic solution
of OJJE 
is obtained
under assumption of the parameter choice
ensuring the phase-lock property.
The closed form of the phase-lock criterion 
in the form of the constraint imposed on the 
problem parameters
is conjectured.

\section{Overdamped \JJ{} equation against 
reduced
double confluent Heun
equation}\label{s1}

The analytic theory of OJJE (\ref{eq::1}),(\ref{eq::2})
can be based on its reduction to the following
system of two {\em linear\/} first order
ODEs \cite{R11a}
\begin{eqnarray}\label{eq::3}
4\imi\omega z^2 \;x'(z)
&=&2z x(z)
+\left[2B z+A\left(z^2+1\right)\right]  y(z),
                                                \nonumber\\
-4\imi\omega z^2\; y'(z)
&=&
\left[2B z+A\left(z^2+1\right)\right]  x(z)+ 2z y(z),
\end{eqnarray}
where $z$ is the free complex variable. 
Indeed, on the universal
covering $\Omega_1\simeq\mathbb{R}\ni t $ 
of the unit circle in $\mathbb{C}$,
 i.e.\ for $z=\exp(\imi\omega t)$,
any non-trivial solution of \Eqs(\ref{eq::3}) determines
a solution of OJJE 
in accordance with the
equation
\begin{equation}\label{eq::4}
  \exp(\imi\ph)=\frac{\Re x-\imi \Re y}{\Re x+\imi \Re y}
\end{equation}
($\Re$ denotes the real part)
supplemented with the continuity requirement.
[The fulfillment of OJJE  
follows from a straightforward computation
taking into account the equality $\overline z=z^{-1}$
holding true on $\Omega_1$
which is utilized for the demonstration that, for real $A,B,\omega$,
the functions
$\Re x,\Re y$ also verify \Eqs(\ref{eq::3}) on  $\Omega_1.$]
Conversely, 
any real valued solution $\varphi(t)$ of OJJE 
induces 
through \Eq(\ref{eq::4})
some `initial data'
$x(0),y(0)$ (with arbitrary  norm $(x^2(0)+y^2(0))^{1/2}>0$)
for \Eqs(\ref{eq::3}). Having solved the latter 
on $\Omega_1$
for $x,y$,
one obtains, applying (\ref{eq::4}),
another phase function obeying OJJE. 
It however must coincide,
due to the identical initial value assumed at $t=0$,
with the 
original $\varphi(t)$.
Finally, since 
these 
$x,y$ defined on $\Omega_1$ simultaneously 
obey 
the 
linear ODEs
(\ref{eq::3}) 
with meromorphic coefficients
which have 
the only singular points $z=0$ and $z^{-1}=0$, 
they can be extended, 
integrating
 (\ref{eq::3})
along radial directions, to analytic functions
defined on
the whole $\Omega=\Omega_1\times \mathbb{R}_+$,
the universal covering of the Riemann sphere with the punctured
 poles $z=0$ and  $z^{-1}=0$.

It is worth noting that on $\Omega_1$
the real valued functions
$\xx{} =\xx{}(t)=\Re x(e^{\imi\omega t}),
\yy{}=\yy{}(t)=\Re y(e^{\imi\omega t})$
satisfy the equations 
\begin{equation}\label{eq::5}
2\dex \xx{}/\mathrm{d} t =  \xx{}+q \yy{},\;
-2\dex \yy{}/\mathrm{d} t = q \xx{}+ \yy{}.
\end{equation}
leading, together with (\ref{eq::4}), to the equation
$
({\dex/\mathrm{d} t})log(\xx{}^2+\yy{}^2)=\cos\varphi
$ 
which implies
\begin{equation}\label{eq::5a}
\mathrm{const}_1\,e^{-t}\le
|\xx{}+\imi\yy{}|^2\le \mathrm{const}_2\,e^t
\; \mbox{for}\;t>0.
\end{equation}
Notice that if $x,y\not\equiv 0$ then 
$\mathrm{const}_1,\mathrm{const}_2$ may be assumed to be strictly positive.
We shall refer to these boundings later on.

The  key observation 
enabling one  
to radically simplify the problem of 
description of the space of 
solutions of \Eqs(\ref{eq::3}) is as follows \cite{R11a}.
Let us introduce the analytic function $v(z)$
which satisfies the equation
\begin{equation}
  \label{eq::6}
\left[
 z^2\frac{\dex^2}{ \dex^2 z} 
+\left(
\mu 
(z^2+1)
-n 
z\right)
\frac{\dex}{\dex z} 
+(2\omega)^{-2}
\right]
v =0,
\end{equation}
where
\begin{eqnarray}
  \label{eq::7}
n&=&-\left(\frac{B}{\omega}+1\right),\;
\mu=\frac{A}{2\omega},\;
\end{eqnarray}
 are the constants replacing original $A,B$
which will be used below whenever it proves convenient.
Then
a straightforward calculation
 shows that
the functions $x,y$ determined by the equations
\begin{eqnarray}
  \label{eq::8}
  v{\hphantom'}&=& \hphantom{(2\omega z)^{-1} }
\llap{\mbox{$\imi \:$}}
 z^{
{n+1\over2}
}
\exp
\left(
\half\mu 
\left(-z+z^{-1}\right)
\right)
(x-\imi y),
\\  
\label{eq::9}
v'
&=&(2\omega z)^{-1}
 z^{
{n+1\over2}
}
\exp
\left(
\half\mu 
\left(-z+z^{-1}\right)
\right)
(x+\imi y)
\end{eqnarray}
verify \Eqs(\ref{eq::3}). 
Conversely, 
defining the 
function $v(z)$ through the solution $x,y$ of \Eqs(\ref{eq::3})
in accordance with \Eq(\ref{eq::8}), 
a straightforward computation proves satisfaction of
\Eq(\ref{eq::9}) and, then,  \Eq(\ref{eq::6}) follows.
Thus, \Eqs(\ref{eq::3}) are equivalent to (\ref{eq::6}) and 
\Eqs(\ref{eq::8}),(\ref{eq::9}) represent the corresponding one-to-one
transformation.

\Eq(\ref{eq::6}) coincides,
after appropriate identification of the constant parameters,
with 
\Eq(1.4.40) from Ref.~\cite{R14}.
It
represents therefore a
particular instance of the double confluent Heun 
equation (DCHE) which can be shown to be in our case non-degenerated.

It is also worth reproducing here the {\it canonical form\/}
of the ``generic'' DCHE as it is given in Ref.~\cite{R14} 
(\Eq(4.5.1)). It reads
\begin{equation}\label{eq::10}
z^2\frac{\dex^2y}{\dex z^2}
+(-z^2+c z+t)\frac{\dex y}{\dex z}+(-a z+\lambda)y=0
\end{equation}
where $a,c,t,\lambda$ are some constants. To adjust it to our case,
a single term has to be eliminated
setting $a=0$. For brevity, we shall name this subclass of DCHE's
{\it reduced.}
Besides, some obvious rescaling of the free variable $z$ is 
to be 
carried out.
After these,
the three constant parameters remained
in the resulting equation
correspond to our
constant parameters
$n,\mu,\omega$ (we shall not need and so omit the reproducing of
the concrete form of this transformation).

The general analytic theory of DCHE is given
in the chapter 8 of the treatise 
\cite{R13}.
DCHE solutions are there represented,
up to nonzero factors given in explicit form,
through the 
Laurent series whose coefficients are assumed to be
computable through the 
`endless' chain  of
3-term linear homogeneous equations (`recurrence relations').
In the present work, we derive the solution of reduced DCHE in a cognate
but more
explicit form.

As a technical limitation,
we also  stipulate
in the present work for the additional
condition to be imposed on the free constant problem parameters 
claiming of them the ensuring of the  phase-lock
property.
On the base of practice of numerical computations,
it can be conjectured that 
such parameter values  fill up a non-empty open subset 
({\em phase-lock area})
in the whole
parameter space (see also the Conjecture A below).
The case where the parameters 
belong to its complement 
is left beyond the scope of the present work.

\section{Formal solution of reduced DCHE by
Laurent series}

Let us introduce  yet another unknown function $E(z)$ replacing $v(z)$
by means of the transformation
\begin{equation}
  \label{eq::11}
  v(z)=
 z^{{n+\pp\over2}-\imi{\mykappa}} 
e^{-{\mu}z}
E(z),
\end{equation}
where the discrete {\em parity parameter $\pp$\/}
may assume
one of the two values, either $\pp=0$ or $\pp=1$,
and
$\mykappa$ 
is some {\em real\/} positive constant
which will be determined latter on.
For $v$ obeying (\ref{eq::6}),
$E(z)$ verifies the equation 
\begin{eqnarray}
\label{eq::12}
0&=&z^3 E''
+z\left[ \left(\pp- 2\imi{\mykappa}\right) z 
- \mu \left(z^2-1\right) \right] E'
\nonumber\\&&
+\left[ \mu 
\left( {n-\pp\over2}+\imi {\mykappa}\right) z^2
\right.
\nonumber\\&&\hphantom{+}
+\left(
(1-\pp)\left({1\over 4} +   \imi {\mykappa}\right) 
-\mykappa^2
-\left({n+1\over2}\right)^2 +\lambda\right) z
\nonumber\\&&\hphantom{+}
\left.
+\mu\left({n+\pp\over 2}-\imi {\mykappa}\right)
\right]E,
 \\&&
\mbox{where}\;\lambda=(2\omega)^{-2}-\mu^2.\nonumber
\end{eqnarray}
Conversely, (\ref{eq::12}) implies the fulfillment of \Eq(\ref{eq::6}).

At first glance, \Eq(\ref{eq::12}) seems `much worse' 
than the original 
DCHE representation.
Nevertheless, 
it is this equation which 
we shall
attempt to solve 
searching  for its solution in the form of  Laurent series
\begin{eqnarray}
  \label{eq::13}
  E=\sum^\infty_{k=-\infty} a_k z^{k}
\end{eqnarray}
`centered' in the points $z=0$ and $z^{-1}=0$
(which are the only singular points for \Eq(\ref{eq::12}))
with unknown $z$-independent coefficients $a_k$.
Then,
carrying out 
straightforward substitution, one gets a sequence of 3-term 
recurrence relations 
binding triplets of neighboring series coefficients 
which can be written down 
either as
\begin{eqnarray}
0&=&
- \mu \left(k-1-{n-\pp\over2}- \imi {\mykappa}\right) a_{k-1}
\nonumber\\&&
+(Z_{k}+\lambda)
 a_k 
+ \mu \left(k+1+{n+\pp\over2} - \imi {\mykappa}\right)
a_{k+1},
\label{eq::14}
\\&&\mbox{where}\;
Z_k=
\left(k+{\pp-1\over2}
-\imi {\mykappa}\right)^2-\left({n+1\over2}\right)^2,
\label{eq::15}
\end{eqnarray}
or as
\begin{eqnarray}
0&=&
- \mu \left(k-1-{n+\pp\over2}+\imi {\mykappa}\right)  
a_{-(k-1)}
\nonumber\\&&
+(\tilde {Z_{k}}+\lambda)
 a_{-k}
+ \mu \left(k+1+{n-\pp\over2} +\imi {\mykappa}\right) 
a_{-(k+1)}, 
\label{eq::16}
\\&&
\mbox{where}\;
\tilde Z_k=\left(k+{1-\pp\over2}+
\imi {\mykappa}\right)^2-\left({n+1\over2}\right)^2
 \label{eq::17}
\end{eqnarray}
and $k=0,\pm1,\pm2\dots$.
The sets of
\Eqs(\ref{eq::14}) and (\ref{eq::16}) are exactly equivalent
and each of them covers  the whole set of equations
the coefficients $a_k$ have to obey.
However, in the approach utilized in the present work,
we shall consider them both in conjunction, employing (\ref{eq::14})
for coefficients $a_k$ with $k\ge -1$ and (\ref{eq::16})
for $a_k$ with $k\le 1$. 
Thus, 
\Eqs(\ref{eq::14}) and (\ref{eq::16}) 
will be considered separately but on the common index 
variation
`half-interval'
$k\ge 0$ 
(remaining legitimate, in principle, for arbitrary integer $k$). 
Obviously, these two equation sets
cover the complete set of conditions imposed to the coefficients $a_k$
and 
are `almost disjoined' intersecting in their 
`boundary'
$k=\pm0$-members alone.

Let us further 
consider for  $k\ge 0$ the following {\it formal infinite 
products\/}
\begin{eqnarray}
  \label{eq::18}
  R_{k}=\prod_{j=k}^\infty M_j,\;
\tilde R_{k}=\prod_{j=k}^\infty \tilde   M_j
\end{eqnarray}
of the $2\times2$ matrices
\begin{eqnarray}
  \label{eq::19}
   M_j&=&
\left(
\begin{array}{ll}
1+\lambda/ Z_{j} & \mu^2/ Z_{j}
\\
1 & 0
\end{array}
\right),
\\\label{eq::20}
\tilde  M_j&=&
\left(
\begin{array}{ll}
1+\lambda/\tilde  Z_{j} & \mu^2/\tilde  Z_{j}
\\
{\tilde Z_{j-1}}/{\tilde Z_{j}} 
& 0
\end{array}
\right).
\end{eqnarray}
It is assumed throughout that 
the matrices $M_j,\tilde M_j$ with {\it larger\/} indices $j$
are situated in products {\it to the right\/} with respect to
ones labeled with {\it lesser\/} index values.

Notice that in the case $\mykappa=0$ and integer $n$,  zero may 
appear in denominators 
of $M$-factors,
making the above definitions meaningless.
This apparent fault admits a simple
resolution 
(see \Eq(\ref{eq::34}) and the discussion following it).
For a while, we temporary leave out
consideration of
such specific parameter choices.

It is also worth 
noting that the above definitions of $R_{k},\,\tilde R_{k}$
may be understood as a concise 
form of representation of
the `descending' recurrence relations
\begin{eqnarray}\label{eq::21}
  R_k&=&M_{k} R_{k+1},\\\label{eq::21a}
 \tilde R_k&=&\tilde M_{k} \tilde R_{k+1},\;k=0,1,\dots
\end{eqnarray}
among the neighboring $R$'s.
These are the only dependencies which 
will be actually
used below in derivations involving $R_{k},\tilde R_k$.

The formulas
(\ref{eq::18})  are `formal' since  
neither the issue of the convergence of such sequences nor
how one has to understand the
`initial values'
 $R_\infty,
\tilde R_\infty$ necessary for the actual 
determination of the `finite index value' $R$-matrices 
are here addressed.

Now
a straightforward calculation 
applying \Eqs(\ref{eq::21}),(\ref{eq::21a})
shows that
the following formulas
\begin{eqnarray}
  \label{eq::22}
\tilde  a_k
&=&
\mu^k 
{
\Gamma\left(1+{n+\pp\over2}- \imi {\mykappa}\right)
\over
\Gamma\left(k+1+{n+\pp\over2}- \imi {\mykappa}\right)
}
(0,1)\cdot R_k \cdot
\left(
\begin{array}{l}1\\0\end{array}
\right)
\\
  \label{eq::23}
\tilde  a_{-k}
&=&
{\mu^k \over {\tilde Z_{k-1}}}
{
\Gamma\left(1+{n-\pp\over2}+ \imi {\mykappa}\right)
\over
\Gamma\left(k+1+{n-\pp\over2}+ \imi {\mykappa}\right)
}
(0,1)\cdot \tilde R_k \cdot
\left(
\begin{array}{l}1\\0\end{array}
\right)
\end{eqnarray}
yield {\it the formal solutions\/}
to \Eqs(\ref{eq::14}) and (\ref{eq::16}),
respectively.

\section{Validation
of the formal solution}

In this section we
show that 
the formal solution of \Eq(\ref{eq::12})
presented in the form of expansion (\ref{eq::13})
with coefficients 
given by \Eqs(\ref{eq::22}),(\ref{eq::23})
represents a well defined analytic function of $z$.
This means, 
first of all,
that the infinite matrix
products $R_k, \tilde R_k$ it involves converge.
Moreover, the convergence takes place 
for any constant parameter values.

The key auxiliary result 
which may be utilized for the proof of
this assertion is as follows. \\[1.ex]
{\bf Lemma.}\\[-4.ex]
\begin{quote}
Let us consider the sequences of complex numbers 
$\alpha_j,\beta_j,\gamma_j,\delta_j$ satisfying 
the following `ascending'
 matrix recurrence relation
\begin{eqnarray}
  \label{eq::24}
\left(
  \begin{array}{ll}
\alpha_j &\beta_j\\
\gamma_j & \delta_j
\end{array}
\right)
&=&
\left(
  \begin{array}{ll}
\alpha_{j-1} &\beta_{j-1}\\
\gamma_{j-1} & \delta_{j-1}
\end{array}
\right)
\left(
\begin{array}{ll}
1+\lambda/ Z_{j} & \mu^2/ Z_{j}
\\
\sigma_j & 0
\end{array}
\right),
\end{eqnarray}
where 
\begin{equation}
  \label{eq::25}
\begin{array}{rcl}
\mbox{either}\;
  \sigma_j&=&1
\\  \mbox{or}\;
  \sigma_j&=&Z_{j-1}/ Z_{j}.
\end{array}
\end{equation}
Then all they converge as $j\rightarrow\infty$. Moreover,
$\lim\beta_j=0=\lim\delta_j$ whereas for $\alpha,\gamma$-sequences there exist 
positive quantities $N_\alpha$, $N_\gamma$
and 
positive integer $j_0$, all depending at most on 
$n,\mu,\lambda,\mykappa,\pp$,
  such that for all  $j> j_0$
\begin{eqnarray}
  \label{eq::26}
  |\alpha_{j}-lim_{j'\rightarrow\infty} \alpha_{j'}|&<&
{
N_\alpha
\max_{j'>j_0}(|\alpha_{j'}|) 
 j^{-1}
},\\
  |\gamma_{j}-lim_{j'\rightarrow\infty} \gamma_{j'}|&<&
{
N_\gamma
\max_{j'>j_0}(|\gamma_{j'}|) 
 j^{-1}
},
\nonumber
\end{eqnarray}
where the maxima are finite.
\end{quote}
The outline of the  Lemma proof can be found in the Appendix.

\smallskip
\noindent
{\it Remark}: 
Formally,
we need not include  in the lemma stipulation the requirement that 
either $\mykappa\not =0$ or $n$ is non-integer (which would a priori evade
possibility of 
contributions with zero $Z_{*}$ in  
denominator) because with fixed constant parameters 
and sufficiently large $j_0$ 
no zero $Z_{*}$ may appear.


Let us return to \Eqs(\ref{eq::18}) and consider the 
four
double-indexed sets of complex numbers
$\{\alpha,\beta,\gamma,\delta\}^{(j_0)}_j$ 
defined as follows:
\begin{eqnarray}
 \label{eq::26a}
\left(
  \begin{array}{ll}
\alpha_{j}^{(j_0)} &\beta_{j}^{(j_0)}\\
\gamma_{j}^{(j_0)} & \delta_{j}^{(j_0)}
\end{array}\right)=R_{j_0}^{(j)}
=\prod_{k=j_0}^{j} M_k.
\end{eqnarray}
It is straightforward to verify that the
sequences obtained by the picking out the elements with common 
value of the upper index
$j_0$
obey
the recurrence relations (\ref{eq::24}) 
for the upper choice in
(\ref{eq::25}). Hence it follows from the Lemma that 
all they converge.
We denote the corresponding limits as $\alpha^{(j_0)}$ etc.
 We have therefore the consistent definition for the infinite matrix
 products
(\ref{eq::18})
\begin{eqnarray}
 \label{eq::26b}
R_{j_0}=
\left(
  \begin{array}{ll}
\alpha^{(j_0)} &\beta^{(j_0)}\\
\gamma^{(j_0)} & \delta^{(j_0)}
\end{array}\right).
\end{eqnarray}
Let us further note that, increasing $j_0$, the `starting' sequence elements
($\alpha_{j_0}^{(j_0)},\alpha_{j_0+1}^{(j_0)}$ for $\alpha$ etc) tend
to the 
corresponding elements of the
idempotent matrix
\begin{eqnarray}
 \label{eq::26c}
M_{\infty}=
\left(
  \begin{array}{ll}1&0\\1&0
\end{array}\right),\;M_{\infty}^2=M_{\infty},
\end{eqnarray}
the discrepancy decreasing  as
$O(j_{0}^{-2})$. On the other hand, 
in accordance with (\ref{eq::26}) the elements of 
$R_{j_0}^{(j)}$ differ from their $j$-limits constituting $R_{j_0}$
by $O(j_{0}^{-1})$-order quantities. This means that
$R_{j_0}$ tends to $M_{\infty}$ as $j_0$ goes to
infinity with the difference going to zero as $O(j_{0}^{-1})$.
In other words, we have the 
\\[1ex]
{\bf Lemma corollary:}\\[-4.ex]
\begin{quote}
1. $R_{j_0}-M_{\infty}=O(j_{0}^{-1})$.\\
\end{quote}
\vspace{-4.ex}
Besides, in view of the convergence,\\[-4.ex]
\begin{quote}
2. The modules of the elements of all the matrices $R_{j_0}$
are bounded from above in total
\end{quote}
--- provided `no-zeroes-in~denominators' condition 
for the parameter choice
is respected, of course.

In according with the above properties,
decomposing $R_{j}$ into the product of the two factors,
 $R_{j}=R_j^{j_0}\cdot R_{j_0}$,
and approximating $R_{j_0} $ by $M_\infty$,
one obtains a simple but important algorithm of computation
of the products (\ref{eq::18}). 
The approximation
\begin{eqnarray}
  \label{eq::18a}
  R_{j}\approx\prod_{k=j}^{j_0} M_k\cdot M_\infty,\;
\end{eqnarray}
is the better, the larger $j_0>j$ is selected.
In the limit, one gets
\begin{equation}
  \label{eq::28a}
  R_j=\left(
  \begin{array}{ll}
\alpha^{(j)} &0\\
\gamma^{(j)} &0
\end{array}
\right).
\end{equation}
This interpretation resolves 
(in a quite obvious way, though)
the aforementioned problem of specification of the
`initial value'
for the sequence $R_j$ 
treated through the `descending' recurrence
relation 
`$[R_\infty]\dots\rightarrow R_j\rightarrow R_{j-1}\rightarrow\dots$'
implied by \Eq(\ref{eq::21}).
(As a matter of fact, an arbitrary constant matrix 
whose product with $M_\infty$ is nonzero, including the unit matrix,
might be used instead of $M_\infty$ in (\ref{eq::18a}), 
affecting  only  the overall normalization of the result
and the accuracy of (\ref{eq::18a})-like approximation for given $j_0$.)

Having introduced the consistent representation of 
the matrices (\ref{eq::18}), it is straightforward 
to do the same for the matrices $\tilde R_j$ (\ref{eq::19}).
The above speculation applies to them with minor modifications as well.
The only distinction is the making use of the
second (lower) choice in \Eq(\ref{eq::25}) 
and the operating with complex conjugated quantities 
(equivalent in our case
to the replacing
$\mykappa$ by $-\mykappa$) throughout.
We shall mark 
the elements of $\tilde R_j$ obtained in this way
with {\em tildes\/} over 
the corresponding $\alpha$'s and $\gamma$'s.

Now, with the consistent interpretation of the products $R_j,\tilde R_j$
(\ref{eq::18})
in hand,
one is able to
calculate the coefficients $a_k,a_{-k}$ for $k=0,1,2\dots$
in accordance with \Eqs(\ref{eq::22})  (\ref{eq::23}).
The triple matrix products 
reduce to 
separate elements of $R$'s (or $\tilde R$'s)
denoted above as  $\gamma^{(k)}$ (for \Eq(\ref{eq::22})),
and $\tilde\gamma^{(k)}$(for \Eq(\ref{eq::23}), respectively)
which are the functions of the parameters $n,\mu,\lambda,\mykappa,\pp$.
Therefore, the sequences
\begin{eqnarray}
  \label{eq::31}
\tilde  a_k
&=&
\mu^k \gamma^{(k)}
{
\Gamma\left(1+{n+\pp\over2}- \imi {\mykappa}\right)
\over
\Gamma\left(k+1+{n+\pp\over2}- \imi {\mykappa}\right)
} 
     \\
  \label{eq::32}
\tilde  a_{-k}
&=&
{\mu^k \tilde \gamma^{(k)}
\over {\tilde Z_{k-1}}}
{
\Gamma\left(1+{n-\pp\over2}+ \imi {\mykappa}\right)
\over
\Gamma\left(k+1+{n-\pp\over2}+ \imi {\mykappa}\right)
}
,\;k=0,1,2,\dots
\end{eqnarray}
are well defined and
solve \Eqs(\ref{eq::14}),(\ref{eq::16}), respectively,
for $k=1,2,\dots$ 

The important feature of the expressions (\ref{eq::31}),(\ref{eq::32})
which is used below
is their
asymptotic behavior for 
large values of the index
$k$ which is easy to derive in explicit
form.
Specifically,
in accordance with inequalities
(\ref{eq::26}),
the set of modules of
$\gamma$- and $\tilde\gamma$-factors
involved 
in \Eqs(\ref{eq::31}),(\ref{eq::32})
is bounded in total from above and 
each of these sequences 
converge to a finite limit.
These imply in particular the validity of the estimates
\begin{equation}
  \label{eq::31a}
|\tilde a_k|  \propto  {\zeta_1^k \over k!},\;
|\tilde a_{-k}|  \propto  {\zeta_2^k \over k^2 k!},
\end{equation}
asymptotically, in the leading order,
for some $k$-independent 
$\zeta_1,\zeta_2$.

\section{Matching condition and phase-lock criterion}

By construction, the
$\tilde a$-coefficients
defined by 
\Eqs(\ref{eq::22}) and (\ref{eq::23})
obey the linear homogeneous equations 
(\ref{eq::14}) and (\ref{eq::16}), respectively,
which are `the same', essentially, differing only
in the associated intervals of the variation of the index,
consisting of the 
positive integers for the former and negative ones for the latter.
However, these two sequences 
cannot be joined, automatically, since they 
are `differently normalized'.
This means, in particular, that their `edge elements'
indexed with zeroes,
generally speaking, differ.
We will denote
them as $\tilde a_0$ and $\tilde a_{-0}$, respectively, 
distinguishing here, in notations, the index `$0$' from the index `$-0$'.

Now, referring to \Eqs(\ref{eq::31}),(\ref{eq::32}),
one notes that 
in view of the factors of  $\Gamma$-functions present in denominators
and
leading to asymptotic behaviors (\ref{eq::31a})
the
following power series in $z$
\begin{equation}
  \label{eq::33}
  E_+(z)=1+\sum_{k=1}^\infty {\tilde a_k\over\tilde a_0}z^k,\;
  E_-(z)=1+\sum_{k=1}^\infty {\tilde a_{-k}\over\tilde a_{-0}}z^{-k},
\end{equation}
admit absolutely converging majorants.
(We assumed above $\tilde a_{\pm0}\not=0$.
Otherwise, i.e.\ if  $\tilde a_{\pm0}=0$,
$\tilde a_{\pm1}$ may not vanish
and the series with the coefficients $\tilde a_{\pm k}/\tilde a_{\pm1}$
can be utilized instead.)
Indeed,
the  Maclaurin series for the exponent can play this role.
Therefore, the series $E_+(z)$ and $E_-(z^{-1})$ define some {\it entire
  functions}
of $z$.
As a consequence, the expression 
\begin{equation}
  \label{eq::34}
E(z)=
{{4\over\pi^2}\sin\left({\pi\over2}(n+\pp- 2\imi {\mykappa})\right)
\sin\left({\pi\over2}(n+\pp+ 2\imi {\mykappa})\right)
\over
(n+\pp- 2\imi {\mykappa})
(n+2-\pp+2\imi {\mykappa})
}
(E_+(z)+E_-(z)-1)
\end{equation}
represents a single-valued function 
analytic everywhere on the Riemann sphere except the poles $z=0,z^{-1}=0$.
They  are the essential singular points for $E$.

It has to be noted that the additional 
$z$-independent fractional factor in (\ref{eq::34})
may be regarded as a specific
common
`normalization' of the  
(\ref{eq::33})-type series which may be, in principle, arbitrary.
However, its given form is,
 essentially, unique as being
fixed (up to 
an insignificant nonzero numerical factor) in view of the following reasons.

The two sine-factors in the numerator 
regarded as holomorphic functions of $n+\pp\pm 2\imi \mykappa$
are introduced for the canceling out 
zeroes in denominators arising 
due to the poles in the factors $Z_{j}^{-1}$ and $\tilde Z_{j}^{-1}$
involved in the products 
(\ref{eq::21}) and regarded as the functions of the same parameters.
%
The set of these (vicious, essentially) singularities
constitute a homogeneous grid which is just covered by the grid of 
roots of the sine-factors  in  (\ref{eq::34}) --- with the two exceptions.
These two `superfluous' sine-factor roots are, in turn, `neutralized'
by the two linear factors in the denominator in (\ref{eq::34}) 
which are therefore also uniquely determined.
As the result, 
in vicinity of 
any zero in denominators in coefficients
of the power series defining 
 $E(z)$ considered as the function of $n+\pp\pm 2\imi \mykappa$
(a root of some $Z_{*}$ or $\tilde Z_{*}$ ),
 the resulting expression takes the form of 
the ratio $\sin x/x\;(x\simeq 0)$  and
is
not now associated with any irregular behavior.
Thus,
as a matter of fact, 
the fractional factor involved in (\ref{eq::34})
is distinguished (up to a numerical factor)
by the claims (i) to cancel out the poles in the
original expressions of the $\tilde a$-coefficients 
(\ref{eq::31}), (\ref{eq::32}) considered as the functions of 
$n+\pp\pm 2\imi \mykappa$
 and
(ii) to introduce neither more roots nor more poles
as a result of
such a `renormalization'.

Now, when plugging the function (\ref{eq::34})
in \Eq(\ref{eq::12})
in order to verify its fulfillment, 
we may provisionally drop out, sparing the space,
$z$-independent fractional factor (restoring it afterwards).

It is important to emphasize that 
the expressions (\ref{eq::31}),
(\ref{eq::32}), by construction,
verify all the 3-term 
recurrence relations (\ref{eq::14}),(\ref{eq::16})
which bind the $a$-coefficients with indices 
{\em of a common sign}, either
non-negative or non-positive.
The only equation
which does not fall into the above categories, 
and, accordingly,
has not been automatically fulfilled,
is the `central'
one binding the coefficients $a_{-1},a_0=1=a_{-0},a_1$, i.e.\
the equation
\begin{equation} \label{eq::35}
 \mu \left(1+{n-\pp\over2}+\imi {\mykappa}\right) a_{-1}
+(Z_{0}+\lambda)
 a_0 
+ \mu \left(1+{n+\pp\over2}-\imi {\mykappa}\right)
a_{1}=0. 
\end{equation}
With normalization adopted in 
(\ref{eq::33}), one has $a_0=1, a_1=\tilde a_1/\tilde a_0,
a_{-1}=\tilde a_{-1}/\tilde a_{-0}$.
Further,
in accordance with (\ref{eq::31}),(\ref{eq::32})
\begin{eqnarray}
  \label{eq::36}
  \tilde a_0&=&\gamma^{(0)},
\tilde a_1=
{\mu\over 1+{n+\pp\over2}-\imi {\mykappa}}\gamma^{(1)}
\nonumber\\
  \tilde a_{-0}&=&{\tilde\gamma^{(0)}\over\tilde Z_{-1}},
\tilde a_{-1}=
{\mu\over 1+{n-\pp\over2}+\imi
  {\mykappa}}{\tilde\gamma^{(1)}\over\tilde Z_0}
\nonumber
\end{eqnarray}
Besides, one has $\gamma^{(0)}=\alpha^{(1)} $,
$\tilde\gamma^{(0)}=
\tilde\alpha^{(1)}\tilde Z_{-1}/\tilde Z_{0} $.
Combining these dependencies,
the following representation of \Eq(\ref{eq::35}) arises
\begin{eqnarray} 
\label{eq::37}
0&=&
 \mu^2{\gamma^{(1)} \over\alpha^{(1)} }
+(Z_{0}+\lambda)
+ \mu^2
{\tilde\gamma^{(1)}\over \tilde\alpha^{(1)}}.
\end{eqnarray}
Accordingly, it is convenient to
introduce the following function of 
the parameters $\mykappa$ and $n,\lambda,\mu,\pp$ 
\begin{eqnarray}
  \label{eq::38}
\Xi&=&
{{4\over\pi^2}
\sin\left({\pi\over2}(n+\pp- 2\imi {\mykappa})\right)
\sin\left({\pi\over2}(n+\pp+ 2\imi {\mykappa})\right)
\over
(n+\pp- 2\imi {\mykappa})
(n+2-\pp+2\imi {\mykappa})
}\times
\nonumber\\&&
\left(
 \mu^2{\gamma^{(1)}\tilde\alpha^{(1)}}
+(Z_{0}+\lambda)\tilde\alpha^{(1)}\alpha^{(1)}
+ \mu^2
{\tilde\gamma^{(1)}\alpha^{(1)}}
\right)
\end{eqnarray}
where $Z_0=
\left({(\pp-1)/2}
-\imi {\mykappa}\right)^2-\left({(n+1)/2}\right)^2
$ (see (\ref{eq::15}))
and $\alpha$'s, $\gamma$'s
are defined as the elements of the convergent matrix products
as follows:
\begin{equation}
  \label{eq::39}
  \left(
  \begin{array}{ll}
\alpha^{(1)}
 &0\\
\gamma^{(1)} &0
\end{array}
\right)
=
\prod_{j=1}^\infty M_j,
\;
  \left(
  \begin{array}{ll}
\tilde \alpha^{(1)}
 &0\\
\tilde\gamma^{(1)} &0
\end{array}
\right)
=
\prod_{j=1}^\infty \tilde M_j
\end{equation}
(see \Eqs(\ref{eq::18})-(\ref{eq::20})).
The fractional multiplier in the first line of \Eq(\ref{eq::38})
coincides with the one entering  \Eq(\ref{eq::34}) and plays the
identical role:
it
eliminates the vicious singularities
arising for specific values of the parameters $n,\mykappa$.
We shall name $\Xi=\Xi(\mykappa,n,\mu,\lambda,\mykappa,\pp)\equiv\Xi(\mykappa;\dots)$
 the {\em discriminant function\/} for brevity. The following
statement holds true.

\smallskip
\noindent
{\bf Proposition}\nopagebreak \\[-3.ex]\nopagebreak
\begin{quote}
Restricting $\mykappa$ to real values,
the equality
$\Xi(\mykappa;\dots)=0$ 
is the necessary and sufficient condition 
for the single valued analytic function (\ref{eq::34})
to verify
\Eq(\ref{eq::12})
everywhere on the Riemann sphere except its poles $z=0,z^{-1}=0$.
\end{quote}

Indeed, the vanishing of $\Xi$ implies the fulfillment 
of (\ref{eq::35}) (where $a$'s are expressed through $\tilde a$'s),
the last equation binding coefficients of the expansion 
(\ref{eq::13}) which has not been fulfilled
as the result of the very coefficients definition. 
Now all
the 3-term recurrence relations
for $a$-coefficients,
 which \Eq(\ref{eq::12}) is equivalent to,
are satisfied and the analytic
function (\ref{eq::34}) verifies \Eq(\ref{eq::12}) everywhere 
except of its own singular points $z=0$ and $z^{-1}=0$.

The equation 
\begin{eqnarray}
  \label{eq::40}
\Xi(\mykappa;\dots) 
=0
\end{eqnarray}
referred to in the above proposition
can be named {\it the matching condition\/} since it enforces
the sequences of the coefficients $a_k$, 
$a_{-k}$, separately obeying the corresponding `halves'
of the equation chain (\ref{eq::14}) (equivalently, (\ref{eq::16}))
to be `matched'
in their `edge' elements $a_{\pm0}=1,a_{\pm1}$.
%

It is worth emphasizing that,
up to this point, the parameter $\mykappa$ (absent
in \Eq(\ref{eq::6}))
has not been restricted in
any way (it was only assumed to be real).
Now and in what follows 
we regard the condition (\ref{eq::40}) 
as $\mykappa$ definition eliminating
this odd `degree of freedom'.
Now it is a well defined function of the other parameters.
It seems interesting enough that
the addition of unspecified constant 
$\mykappa$ to the transformation (\ref{eq::11}) and its subsequent 
`fine tuning' by means of the claim of fulfillment of
\Eq(\ref{eq::40}) is necessary 
for the representation of 
solution of DCHE (\ref{eq::6}) in terms of convergent Laurent series.
More precisely, it is clear that \Eqs(\ref{eq::15}) can be solved
{\em for any\/} (including trivial zero) choice
of $\mykappa$,
choosing loosely  $a_{j_0},a_{j_0+1}$ for 
arbitrary $j_0$ and then
calculating, term by term, all the coefficients
$a_j$, advancing in parallel in both
directions of $j$-index variation `from $j_0$ towards $\pm\infty$'. 
Then (\ref{eq::13}) immediately yields a ($\mykappa$-dependent!) 
formal solution to \Eq(\ref{eq::12}) and hence, 
through transformation (\ref{eq::11}), 
to \Eq(\ref{eq::6}).
However, it 
can be only formal and
will necessarily {\em  diverge\/} for any $z$ unless 
the matching condition (\ref{eq::40}) is fulfilled --- just in view of the
uniqueness of solution with the analytic properties presupposed.
On the other hand,
considering separately the `halves' of the set of \Eqs(\ref{eq::15})
and resolving them `in index variation directions' 
opposite to the ones assumed above 
(in a sense, `from $\pm\infty$ towards $\pm0$'), 
we obtain the always {\em converging\/}
series (\ref{eq::33}). However, as we have seen, 
we again have no solution (in this case even formal)
unless the matching condition 
fixing $\mykappa$
is fulfilled. Obviously, the uniqueness property
implies that the introduction of the 
`branching'
power function factor, 
as in 
(\ref{eq::11}), is the only way to obtain a solution to \Eq(\ref{eq::6})
admitting representation in terms of convergent  power series.

Now,
tracking back the relationship connecting \Eq(\ref{eq::12})
with the primary~\Eq(\ref{eq::1})
and invoking the general theory of the latter 
applicable in the case of arbitrary continuous periodic $q(t)$ \cite{R11},
one can infer the following statements
which however, in the present context, are only of 
the status of {\em conjectures\/} in view of the lack of
their proof `from the first principles',
i.e.\ on the base of the properties of the discriminant function $\Xi$
following from its definition.

\smallskip
\noindent{\bf Conjecture A.}\\[-4.ex]
\begin{quote}
1. There exists an open non-empty subset of the space of the
problem parameters $n,\mu,\omega,\pp$ where 
\Eq(\ref{eq::40})
admits a real valued positive solution.\\
2. If real $\mykappa$ solves \Eq(\ref{eq::40}), $-\mykappa$
is the solution as well. No more real roots exist.\\
3. Real roots of \Eq(\ref{eq::40}) obey the condition
$|\mykappa|\le (2\omega)^{-1}$.
\end{quote}
\noindent
{\it Remark}: 
The last statement is nothing else
but the form of the limitation on the rate of
growth or decreasing of the functions $\xx{},\tilde
y=x,y|_{z=e^{\imi\omega t}}$ of the real variable $t$
implied by the inequalities (\ref{eq::5a}) and the equation
$$
x-\imi y=-iz^{{\pp-1\over2}-\imi \mykappa}
e^{-{\mu\over2}\left(z+{1\over z}\right)} E,
$$
following from definitions.
Notice that the latter clarifies the role of the {\em parity\/} 
parameter $\pp$ which 
determines the multiplicity of the inverse to the map 
$\mathbb{S}^1\ni z\rightarrow (x+\imi y)/|x+\imi y|\in\mathbb{S}^1$ induced
by the solution (\ref{eq::34}). 
If $\pp=0$, the revolution along the circle $|z|=1$
leads to the reversing of the direction of the vector
with components
$(x,y)$ whereas in the case $\pp=1$ its direction is preserved.
The inverse map is double-valued in the former case and one-to-one
in the latter one.

More properties of the discriminant function
can be inferred from the 
numerical experiments although, as opposed to the assertions 
of the Conjecture A, they
have, to date, no analytic arguments in their favor yet, even indirect. 
Nevertheless, the first item below is 
important enough from viewpoint of applications 
(seeming also plausible enough)
to be
{\em explicitly formulated\/} here.\\[1.ex]
{\bf Conjecture B.}\\[-4.ex]
\begin{quote}
1. {\bf Phase-lock criterion.}\\
 Equation  (\ref{eq::40}) admits
a real non-zero solution if and only 
if 
\begin{equation}\Xi(0;\dots)>0.\end{equation}
\end{quote}
\vspace{-1.ex}
This means in particular that $\Xi(0;\dots)$ is real; moreover,
the numerical study
makes evidence that\\[-4.ex]
\begin{quote}
2. $\Xi(\mykappa;\dots)$ is  real for real $\mykappa$ 
\end{quote}
 (assuming the other
parameters to be also real, of course).

\section{Floquet solutions of DCHE and  involutive solution maps
 }

Let us assume now that there exists a real positive solution $\mykappa$ of
\Eq(\ref{eq::40}). With this $\mykappa$,
the function $E(z)$
defined by \Eq(\ref{eq::34})
 verifies \Eq(\ref{eq::12}). 
Let us 
consider the function $\Ep(z)$ defined through  $E(z)$
as follows:
\begin{eqnarray}
  \label{eq::41}
  \Ep(z)&=&z^{2\imi{\mykappa}-\pp}
\left[ E'\left({1\over z}\right)
+\left(\left({n+\pp\over2}-\imi{\mykappa}\right)z-\mu\right)
E\left({1\over z}\right)\right].
\end{eqnarray}
Then a straightforward computation shows that it verifies 
\Eq(\ref{eq::12}), provided $E(z)$ does. 

It is worthwhile to note that the 
repetition of the
transformation (\ref{eq::41}) yields no more solutions to \Eq(\ref{eq::12}).
As a matter of fact, one has $_{\#}\circ_{\#}=(2\omega)^{-2} Id$.
Thus $(2\omega)^{-1}{}_\#$ is the involutive map on the space
of its solutions.

Next, the functions
$E(z)$
and $\Ep(z)$ are linearly independent for nonzero
$\mykappa$. Indeed, utilizing (\ref{eq::12}), one finds
\begin{eqnarray}
  \label{eq::42}
   \dEp(z)&=&z^{ 2\imi{\mykappa}-\pp-1}
 \left[
 \left(-{n+\pp\over2}+\imi{\mykappa} +\mu z\right)
 E'\left({1\over z}\right)
 +
 \right.
 \nonumber\\
 &&\left.\hspace{-14ex}
 +
 \left(
 \mu\left({n+\pp\over2}-\imi{\mykappa}\right)(z^2+1)
 +
 \left( -\left({n+\pp\over2}-\imi{\mykappa}  \right)^2
 +\lambda
 \right)z
 \right) E\left({1\over z}\right)
 \right].
\end{eqnarray}
This reduction allows one 
to calculate
the determinant of the linear transformation 
binding the pairs of functions $\Ep,\dEp(z)$ and $E,E'(1/z)$
which proves
equal to  $(2\omega)^{-2}z^{2(-\pp + 2\imi\mykappa)}$ and is therefore
 nonzero. 
Hence $\Ep(z)$ is not identically zero
(and may vanish at isolated points, at most, as well as
$E(z)$).
Finally,
$E(z)$ is periodic on the unit circle centered at zero whereas $\Ep(z)$,
for real $\mykappa\not=0$,
is not. Hence they are linear independent.
The functions
$E(z)$ and $\Ep(z)$ constitute therefore the fundamental
system for \Eq(\ref{eq::12}) and any its solution 
can be expanded in this basis with constant expansion coefficients.
The analytic properties of these solutions identify  them
as the unique pair of the
{\it Floquet solutions\/} of the reduced DCHE under consideration.
See \cite{R13}, section 2.4.

The function $E(z)$ obeys the important functional equation 
which can be derived as follows. 
A straightforward calculation
shows that the {\sc rhs} expression of \Eq(\ref{eq::41})
with the `branched' factor
$z^{2\imi{\mykappa}}$ removed (i.e.\ 
$z^{-2\imi{\mykappa}} \Ep(z)$)
satisfies the ODE which coincides with (\ref{eq::12})
{\em up to the opposite sign of the parameter $\mykappa$}. This means
that, for real $\mykappa$, $n,\mu,\omega$,
 the analytic function
\begin{eqnarray}
 \label{eq::43}
\hat E(z)&=&
z^{-\pp}
\left[
 \bar E'\left({1\over z}\right)
+\left(\left({n+\pp\over2}+\imi{\mykappa}
\right)z-\mu\right)\bar E\left({1 \over z}\right)
\right],
\end{eqnarray}
where $\bar E(z)=\overline{E(\overline z)}$,
is the solution of  \Eq(\ref{eq::12}) itself.

As opposed to $\Ep$, this `yet another' solution $\hat E(z)$ has 
{\it the same\/} analytic
properties as $E(z)$ and 
hence must coincide with it
up to some numerical factor $C_C$, i.e.\
\begin{equation}
  \label{eq::44}
C_C E(z)=
  z^{-\pp}
\left[
 \bar E'\left({1 \over z}\right)
+\left(\left({n+\pp\over2}+\imi{\mykappa}
\right)z-\mu\right)\bar E\left({1 \over z}\right)
\right].
\end{equation}
($C_C$ may not vanish since otherwise $\Ep$
would also be zero.) This is the generalization of the
similar property of  the so
called `Heun polynomials' established in Ref.~\cite{R11a}.

The complex valued constant $C_C$
actually reduces  to a single real constant.
To show that, let us 
notice at first that if $E(z)$ verifies \Eq(\ref{eq::12}) 
it follows from the latter and (\ref{eq::44})
\begin{eqnarray}
  \label{eq::45}
C_C E'(z)&=&
z^{-\pp-1}
\left[
\left(
  -{n+\pp\over2}-\imi{\mykappa}
+\mu z
\right)\bar E'(1/z)
\right.
\nonumber
\\&&
+
\left(
 \mu    {\pp+n\over2}(z^2+1)
-\imi\mu{\mykappa}(z^2-1)
\right.
\nonumber
\\&&\left.\left.
+\left(
-{(n+\pp)^2\over4}
-{\mykappa^2}+\lambda
\right)z
\right)
\bar E(1/z)
\right].
\end{eqnarray}
Evaluating now \Eqs(\ref{eq::44}),(\ref{eq::45})
together with their complex conjugated
versions with $z=z^{-1}=1$, one 
obtains  four linear homogeneous equations binding 
the quantities $E(1),E'(1),\bar E(1),\bar E'(1)$ which may not
vanish simultaneously. The corresponding consistency condition 
reads 
$ |C_C|^2=(2\omega)^{-2}$ implying
\begin{equation}
  \label{eq::46}
 C_C=(2\omega)^{-1}e^{\imi C_c},
\end{equation}
where $C_c$ is some {\em real\/} constant (actually, the function of the
parameters $n,\mu,\omega,\pp$). It
encodes all the monodromy data for \Eq(\ref{eq::12}), essentially.

It is straightforward to 
show that the transformation 
\Eq(\ref{eq::44})
is also involutive. 
It manifests the specific symmetry
in the behaviors of the function $E(z)$ in vicinities
of the essentially singular points $z=0$ and $z^{-1}=0$.
Remarkably, this symmetry
implies itself the fulfillment
 of \Eq(\ref{eq::12}).
Indeed,
differentiating (\ref{eq::44})
and taking into account (\ref{eq::46}), 
one arrives at \Eq(\ref{eq::12}).
In a sense, \Eq(\ref{eq::44}) 
together with 
stipulation for the  analyticity of $E(z)$
can be considered as the equivalent to \Eq(\ref{eq::12}).
Additionally, \Eq(\ref{eq::44})
implies anti-linear (involving complex conjugation)
dependencies among the `distant' 
Laurent series
coefficients $a_{-k}$ and $a_{k},a_{k+1}$ (whereas
\Eq(\ref{eq::14}) (or (\ref{eq::16}))
binds `nearby' 
$a_k,a_{k\pm1}$).
In particular, it suffices to find all $a_k$ for $k>0$
and then $a_{-k}$ can be computed
from the latter by means of a simple transformation.

\section{Essentially periodic and general
solutions of overdamped Josephson junction equations
}

The connection between the functions $E(z),\Ep(z),\hat E(z)$ 
pointed out  
above is important for the lifting 
the results concerning solutions of \Eq(\ref{eq::12})
to the level of original OJJE. 
This procedure applies \Eqs{}
(\ref{eq::4}),
(\ref{eq::8}), 
(\ref{eq::9}),
(\ref{eq::11}) and leads to the following conclusions.

At first, 
the representation of the two special (and the most important)
solutions to OJJE 
for which the 
exponents $\exp(\imi\varphi)$ are {\it periodic\/} (for brevity, we shall call
such phase functions {\em essentially periodic})
follows. It reads
\begin{eqnarray}
  \label{eq::47}
  \exp(-\imi\varphi)&=&
2\imi\omega
\left(
z {E'(z)\over E(z)}
+{n+\pp\over2}-\imi{\mykappa}-\mu z,
\right),
\\
  \label{eq::48}
  \exp(\imi\varphi)&=&
-2\imi\omega
\left(
z^{-1}{E'(z^{-1})\over E(z^{-1})}
+{n+\pp\over2}-\imi{\mykappa}
-\mu z^{-1}
\right),\\
&&\mbox{where}
 \; z=\exp(\imi\omega t).\nonumber
\end{eqnarray}
For $\mykappa>0$
the first of these formulas 
determines 
the asymptotic 
limit (the attractor) of a generic solution
whereas the second solution
is unstable (the  repeller).
It is important to emphasize that the functions $\varphi(t)$
defined by \Eqs(\ref{eq::47}) and (\ref{eq::48}) are {\em real\/} and
\Eq(\ref{eq::46}) is the crucial property utilized in
the calculation
establishing
this fact.

At second, 
it is straightforward to obtain 
the `nonlinear superposition' of solutions (\ref{eq::47}),
(\ref{eq::48})
operating with their DCHE-related counterparts. 
The result is represented by the formula
\begin{eqnarray}
  \label{eq::49}
\exp(\imi\varphi)&=& 
-{\imi\over2}
\left\{\cos\psi\cdot
E(z)
+\sin\psi\cdot 
z^{-\pp+2\imi{\mykappa}}
\times
\vphantom{\left({n+\pp\over2}-\imi{\mykappa}-\mu z\right) }
\right.
\nonumber\\
&&
\left.
\left[
E'(z^{-1})
+\left(
\left({n+\pp\over2}-\imi{\mykappa}
\right)z
-\mu \right)E(z^{-1})
\right]
\right\}\times
\nonumber\\&&
\left\{
\omega\cos\psi\cdot
\left[
z E'(z)+\left({n+\pp\over2}-\imi{\mykappa}-\mu z\right)E(z)
\right]
\right.
\nonumber\\&&\left.
+
{1
\over4\omega} \sin\psi\cdot
z^{-\pp+1+2\imi{\mykappa}}
E(z^{-1})\right\}^{-1},
\end{eqnarray}
where $\psi$ is an arbitrary real constant.
More exactly, the set of all functions $\varphi$ 
described 
by \Eq(\ref{eq::49})
is parameterized by a point on
the unit circle.
As opposed to 
 (\ref{eq::47}),
(\ref{eq::48}), 
the function (\ref{eq::45}) 
is defined on the universal covering of the Riemann sphere with punctured
poles, $\Omega$.
Continuous (and then real analytic) function $\ph(t)$ determined
by this equation on $\Omega_1\in\Omega$, where $z$ is understood as $e^{\imi\omega t}$,
is just
the general solution of OJJE 
in the case of phase-lock.

In particular,
\Eqs(\ref{eq::47}),(\ref{eq::48}) arise as particular
cases of (\ref{eq::49}) for $\psi=0$ and $\psi=\pi/2$, respectively.
As a consequence, 
asymptotic properties of the general solution mentioned above
immediately follow.
Indeed,
 as $t$ increases, the exponent  (\ref{eq::49})
is converging to 
(\ref{eq::47}) and is moving off 
(\ref{eq::48}) (unless it coincides with the latter).
 The two
solutions described by 
(\ref{eq::47}),(\ref{eq::48}) are the only
ones which are not affected by the translations 
$t\rightarrow{}t+2\pi\omega^{-1}$
(in the sense 
the exponents (\ref{eq::47}),(\ref{eq::48}) are kept unchanged)
and preserve their form in asymptotics.

At third,
considering $\varphi$ defined by (\ref{eq::47})
as analytic function of
$z$  and taking in account \Eq(\ref{eq::12}),  
one obtains
\begin{eqnarray}
  \label{eq::50}
  {\dex\varphi\over\dex z}&=&
-\imi z^{-2} 
\left\{
              z^3 \left({E'(z)\over E(z)}\right)^2
+z\left(\left(1-z^2\right)\mu
                     +z\left(\left(\pp-1\right)
-2\imi{\mykappa}\right)\right)
               {E'(z)\over E(z)}
\right.
\nonumber\\
&&
\left.
-\left(z+z\left({n-\pp\over2}+\imi{\mykappa}\right)-\mu\right)
\left(\left({n+\pp\over2}-\imi{\mykappa}\right)-\mu z\right)
+{z\over4\omega^2}\right\}
\times
\nonumber\\
&&
\left\{
z {E'(z)\over E(z)}
+\left({n+\pp\over2}-\imi{\mykappa}\right)
-\mu
  z
\right\}^{-1}
\end{eqnarray}
On the unit circle, this $\varphi$ verifies OJJE.
It is therefore
smooth (even real analytic). 
Then (\ref{eq::50}) is continuous on the unit circle.
Finally, since $\exp\imi\varphi|_{z=\exp\imi\omega t }$ is 
periodic, the following proposition 
holds true:\\

\smallskip
\noindent
{\bf Proposition 2.}\\[-3.ex]
\begin{quote}
The quantity
\begin{eqnarray}
  \label{eq::51}
  k=(2\pi)^{-1}
\oint%
{\dex\varphi\over\dex z}
\dex z,
\end{eqnarray}
where ${\dex\varphi/\mathrm d\, z}$ denotes the {\sc rhs} 
expression from
\Eq(\ref{eq::50})
and the
integration is carried out over the circle $|z|=1$,
is well defined and equals to an 
integer.
\end{quote}
This integer is  the degree
of the map $\mathbb{S}^1\Rightarrow\mathbb{S}^1$
induced by the function (\ref{eq::47}).
In physical applications, it is called {\it the phase-lock order\/}
and is considered as an integer-valued function of the parameters.
Phase-lock order is involved in the formula 
representing the 
property of being `essentially periodic' for
the phase function defined by \Eq(\ref{eq::47})
(and asymptotically for a generic phase function)
which reads
$$\forall t\;
\ph(t+2\pi\omega^{-1})=\ph(t)+2\pi k
$$
In a phase-lock state of JJ,
the uniformly distributed discrete 
levels of averaged voltage 
equal to 
$k\cdot (\hbar\omega/2e)$ for some $k=0,\pm 1,\pm 2\dots$
are observed.

\smallskip
\noindent{\bf Conjecture C.}\\[-4.ex]\nopagebreak
\begin{quote}
Any integer map degree (\ref{eq::51})
is realized on some
non-empty open subset of the space of the
problem parameters $n,\mu,\omega,\pp$.
\end{quote}
This assertion is closely cognate to the item 1 of the above Conjecture A.

\section{Summary}

It the present work, the general solution of the
overdamped Josephson equation (\ref{eq::1}) is derived
for the (co)sinusoidal {\sc rhs} function (\ref{eq::2})
in the case of one of three possible asymptotic
behaviors known as the phase-lock mode.
The solution is represented in explicit form
in terms of the Floquet solution 
of the particular instance (corresponding to the vanishing
of one of the four free constant parameters)
of the
double confluent Heun equation (DCHE).
The Floquet solution of DCHE is represented in terms of 
the Laurent series whose coefficients are determined by
the convergent
infinite products of $2\times 2$ matrices with a single 
zero element tending to the idempotent matrix (\ref{eq::26c}). 
The derivation 
presupposes the existence
of a real solution of the transcendental
equation
(\ref{eq::40}) 
which  is equivalent
to the claim of realization of the phase-lock mode
for the given parameter values.
The plausible criterion 
of its existence (i.e.\ {\em the phase-lock criterion})
is conjectured.

It is worth summarizing here the main steps of  solution of
OJJE. 
They can be condensed as follows.
\begin{itemize}
\item
The investigation of the basic properties
of \Eq(\ref{eq::1})
for arbitrary periodic (sufficiently regular)
$q(t)$ allows one to establish the division of the
space of the problem parameters 
into the two open areas of which one  corresponds
to the phase-lock property 
of the OJJE  solutions
whereas another corresponds to their pseudo-chaotic
behavior revealing no stable periodicity. 
For the (lower-dimensional) complement to these areas the
intermediate behavior is observed.
The corresponding results are discussed in
sufficient details 
in Ref.~\cite{R11}. 
\item
The next important point 
is the intimate connection
(first mentioned by V.~Buchstaber, see, e.g., 
Ref.~\cite{R12})
between (\ref{eq::1})
and a  simple {\em linear} system 
of the two first order ODEs
(\ref{eq::5}). For (co)sinusoidal {\sc rhs} function (\ref{eq::2}),
the latter
takes the form (\ref{eq::3}).
\item
At the next step, the transformation
(\ref{eq::8}) was found which converts the linear system (\ref{eq::3})
to a
particular instance of the \DCHE{} (\ref{eq::10}).
\end{itemize}
Generally speaking,
it could be solved by means of the expansion in Laurent 
series
\cite{R13},\cite{R11a}
centered at 
the singular points
but preliminarily
the additional simple but important transformation
has to be carried out:
\begin{itemize}
\item{}
the `branched' power factor involving unspecified constant
($\mykappa$-dependent contribution in \Eq(\ref{eq::11}))
is introduced. 
\item{}
The addition of the discrete `parity' parameter $\pp$,
assuming either the value 0 or the value 1, which
is involved in the  power factor
proves necessary for the subsequent exhaustive 
`indexing' of the solution space.
\item{}
After that, the standard technique of the
power expansion leads to the `endless' 
sequence of the 3-term constraints (\ref{eq::14}) 
(or, which is the same, (\ref{eq::16}))
imposed on the 
unknown series coefficients.
\item{}
The next step is the devision 
of the set of power series coefficients into two subsets.
The
non-negative-index-value coefficients and 
non-positive-index-value ones
are treated separately, solving the
separate subsets of the equations (\ref{eq::14}) and (\ref{eq::16}),
respectively, for $k\ge 1$.
The application of 
the continued fraction technique leads, after some transformations,
to the `explicit' formulas 
for the series coefficients
involving infinite products of
$2\times 2$ matrices converging for large index values to the
idempotent matrix (\ref{eq::26c}).
This convergence is sufficiently fast to imply the 
convergence of the matrix products and, accordingly, the finiteness of 
the series coefficients. Moreover,
the associated estimates 
make evident the existence of 
the absolutely converging
majorants for the resulting Laurent series.
Therefore, they actually determine
the Floquet solution of DCHE.
The latter proves 
representable as the sum of the two entire functions of the
arguments $z$ and $z^{-1}$, respectively.
\item 
The procedure
 producing Laurent series coefficients 
noted 
above proves suffering
however from 
the improper 
introduction 
of a kind of
vicious singularities arising as zeroes in denominator
which appear for some special parameter values.
They are eliminated my means of multiplication
of the `raw' coefficient expressions
by some $z$-independent (but parameter dependent)
factors given in explicit form.
\item
Now the `solution candidate' for \Eq(\ref{eq::12})
can be represented as the analytic function (\ref{eq::34})
which is well-defined for any parameter values.
However, at the price of automatic convergence
of the series it has been built upon from,
it does not always verify \Eq(\ref{eq::12}).
The equation is fulfilled if and only if the 
fulfillment 
  of \Eq(\ref{eq::40}), 
which is the transcendental equation
for the still unspecified parameter $\mykappa$,
is stipulated. 
\end{itemize}
At that stage,
having solved \Eq(\ref{eq::40}),
a single 
solution (the Floquet solution) of DCHE can be regarded as having been 
explicitly constructed.
\begin{itemize}
\item
The invariance of the space of solutions of DCHE
under consideration 
with respect to transformation 
represented by
\Eq(\ref{eq::41})
allows one to immediately obtain the fundamental system of 
its solutions in terms of the single Floquet solution
noted above.
\item
The automorphism represented by 
\Eq(\ref{eq::44}) expresses the important intrinsic property 
of the Floquet solution of DCHE.
It is used 
for the derivation of
the explicit representation of 
the exponent $\exp(\imi\varphi)$
specifying the real valued
phase function $\varphi$ to be obtained as 
the restriction 
of the analytic function from the universal covering 
of the Riemann sphere with punctured poles (\Eq(\ref{eq::49}))
to the universal covering of the unit circle.
It yields 
the general solution of 
\Eqs(\ref{eq::1}),(\ref{eq::2}) {\em in the case of phase-lock}.
\item
Employing analytic properties 
of 
$\exp(\imi\varphi)(z)$,
the
formula (\ref{eq::51}) involving Floquet solution of DCHE
follows
which gives the 
degree 
of the map 
$\mathbb{S}^1\Rightarrow\mathbb{S}^1$ 
it induces (the winding number) 
also known in application fields as the phase-lock order.
\end{itemize}

It is worth noting in conclusion that
all the constructions derived above 
admit a straightforward
algorithmic implementation
which have been used for the numeric verification
of the relevant relationships.

\appendix
\section{Appendix: outline of the Lemma proof}


\Eq(\ref{eq::24}) implies
$  \beta_j=\mu^2\alpha_{j-1} Z_{j}^{-1},
  \delta_j=\mu^2\gamma_{j-1} Z_{j}^{-1}$
and hence the asserted properties of the sequences $\beta_j,\delta_j$
follow from 
the existence of finite limits for  the sequences $\alpha_j,\gamma_j$.
As to 
$\alpha_j$ and $\gamma_j$, they have
to obey the identical decoupled 3-term  recurrence relations 
which, for $\alpha$'s, read
\begin{equation}
  \label{eq::52}
  \alpha_j=
(1+\lambda  Z_{j+(\pp-1)/2}^{-1})\alpha_{j-1}
+
\mu^2  Z_{j-\tilde\pp+(\pp-1)/2}^{-1}\alpha_{j-2}
\end{equation}
where   $\tilde\pp=1$ for the upper choice of $\sigma$
in (\ref{eq::25}) and 
$\tilde\pp=0$ for the lower $\sigma$ choice therein.
It suffices to consider the $\alpha$-sequence case.

Evidently, for every integer $j_0>0$ and $l>0$
any solution of equations (\ref{eq::52})
can be represented in terms of the
decomposition
\begin{equation}
  \label{eq::53}
\alpha_{j_0+l}=(1+p_{j_0,l})\alpha_{j_0-1}+q_{j_0,l}\alpha_{j_0-2},
\end{equation}
for some coefficients
$p_{j,l},q_{j,l}$ 
independent on the `starting' terms $\alpha_{j_0-1},\alpha_{j_0-2}$.
Applying (\ref{eq::52}),
it is straightforward to show 
\begin{eqnarray}
  \label{eq:53a}
q_{j_0,l+1}&=&(1+p_{j_0+1,l} )\mu^2 Z_{j_0-\tilde\pp}^{-1},
\end{eqnarray}
whereas for $p_{*,*}$ one gets the following recurrence relation:
\begin{equation}
  \label{eq::54}
  p_{j_0,l+1}=
p_{j_0+1,l}+
\lambda(1+p_{j_0+1,l}) Z_{j}^{-1}
+\mu^2(1+p_{j_0+2,l-1})Z_{j+1-\tilde\pp}^{-1}.
\end{equation}
\Eq(\ref{eq::54}) is equivalent to 
(\ref{eq::52}), essentially, but it possesses 
the advantage of being endowed
with the standard `initial conditions'
\begin{eqnarray}
  \label{eq::55}
p_{j_0,-1}&=&0, p_{j_0,-2}=-1
\end{eqnarray}
which follow from definitions.
Besides, one gets
\begin{eqnarray}
  \label{eq::56}
q_{j_0,-1}&=&0, q_{j_0,-2}=1. 
\end{eqnarray}

It proves convenient to carry out one more rearrangement
of unknowns introducing the differences
\begin{equation}
  \label{eq::57}
\Delta p_{j_0,l}=p_{j_0,l+1}-p_{j_0,l}.
\end{equation}
which obey the own `initial conditions'
\begin{eqnarray}
  \label{eq::58}
\Delta p_{j_0,-2}&=&
1,
\\
 \label{eq::58a}
\Delta p_{j_0,-1}&=&
\lambda Z^{-1}_{j_0},
\end{eqnarray}
and similar  recurrence relations
\begin{equation}
  \label{eq::59}
\Delta  p_{j_0,l+1}=
\Delta p_{j_0+1,l}+
\lambda \Delta p_{j_0+1,l} Z_{j_0}^{-1}
+\mu^2 \Delta p_{j_0+2,l-1}Z_{j_0+1-\tilde\pp}^{-1}.
\end{equation}

Now,
summing up the subset of the latter 
with the common sum of indices 
at the left and taking into account (\ref{eq::58a}),
all but two `free' $\Delta p$-terms cancel out and
one obtains the following 
equation 
\begin{eqnarray}
  \label{eq::60}
 \Delta  p_{j_0,l+1}&=&\lambda Z^{-1}_{j_0+2+l}
 \nonumber\\&&
+\lambda \sum_{m=0}^{l+1}
\Delta p_{j_0+1+m,l-m} Z_{j_0+m}^{-1}
 \nonumber\\&&
+\mu^2 \sum_{m=0}^{l+1}
\Delta p_{j_0+2+m,l-1-m}Z_{j_0+m+1-\tilde\pp}^{-1}.
\end{eqnarray}
In the sums, the second index of $\Delta p_{*,*}$ is 
everywhere less than the same index at the left
that allows to apply the method of mathematical induction.
For the `starting' values -1,0 of the second index one has
\begin{eqnarray*}
 Z^{}_{j_0}\Delta p_{j_0,-1}&=&\lambda,
\nonumber
\\
 Z_{j_0+1}
\Delta p_{j_0,0}&=&
\lambda(1+\lambda Z^{-1}_{j_0})
+\mu^2 Z_{j_0+1} Z^{-1}_{j_0+2-\tilde\pp}.
\end{eqnarray*}
Therefore for $l=-1,0$ there exist the finite limits 
 $\lim_{j_0\rightarrow\infty}|Z^{}_{j_0+l+1}\Delta p_{j_0,l}|$.
As a consequence,
for these $l$'s one has 
\begin{equation}
  \label{eq::61}
  |\Delta p_{j_0,l}|< \tilde N |Z_{j_0+l+1}|^{-1}
\end{equation}
for  appropriate constant $\tilde N$ which is convenient to choose $>1$.
Let us consider this fact as the starting point 
of mathematical induction and assume that for some integer $l_0\ge 0$
and any integer $l$ from the interval $[-1,l_0]$, (\ref{eq::61})
holds true. We may apply it for the estimating from above of the quantity
$|\Delta p_{j_0,l_0+1}|$. This can be realized making use of 
the `decomposition'
(\ref{eq::60})
and the following 
 elementary inequalities
\begin{eqnarray}\label{eq::62}
\sum_{m=j_0}^{L+j_0+1}
|Z_{m}|^{-1}&<&{
1+|n+1|^{-1}
\over
j_0-|n+1|/2+(\pp-1)/2
},
\\
\sum_{m=j_0+1-\tilde\pp}^{L+j_0+2-\tilde\pp}
|Z_{m}|^{-1}&<&
{
1+|n+1|^{-1}
\over
j_0-|n+1|/2+(\pp-1)/2
},
\end{eqnarray}
where  $L>0$ (and $n\not=-1$) .
These imply the inequalities
\begin{eqnarray}
| \Delta  p_{j_0,l_0+1}|
&\le&
|\lambda| |Z_{j_0+l_0+2}|^{-1}
 \nonumber\\&&
+\tilde N
|Z_{j_0+l_0+1}|^{-1}
\left(
 |\lambda| 
\sum_{m=0}^{l_0+1}
|Z_{j_0+m}|^{-1}
+
|\mu^2|
\sum_{m=0}^{l_0+1}
|Z_{j_0+m+1-\tilde\pp}|^{-1}
\right)
\nonumber\\&<&|Z_{j_0+l_0+2}|^{-1}
\left(1+\tilde N{|Z_{j_0+l_0+2}|\over |Z_{j_0+l_0+1}|}
{
(|\lambda|+|\mu^2|)(1+|n+1|^{-1})
\over
(j_0-|n+1|/2+(\pp-1)/2)
}
\right).\nonumber
\end{eqnarray}
Since we assumed $\tilde N>1$,  there exists the  
lower index value bound such that for any $j_0$ exceeding
it
the factor in brackets is less than $\tilde  N $ and then
the above inequalities imply
$|\Delta p_{j_0,l_0+1}|<\tilde  N |Z_{j_0+l_0+2}|^{-1}$.
(\ref{eq::61}) is therefore 
established
for sufficiently large $j_0$ and arbitrary
$l\ge0$. Increasing $\tilde N$ if necessary, (\ref{eq::61}) 
proves valid for arbitrary $j_0$.

In view of this property, one sees that the  
sum $\sum_{l=0}^\infty\Delta p_{j_0,l}$
has the majorant $\sum_l|Z_{j_0+l+1}|^{-1}$ and thus converges itself.
The sequence of its partial sums 
 $\sum_{m=0}^l\Delta p_{j_0,m}=p_{j_0,l+1}-p_{j_0,-1}=p_{j_0,l+1}$
also converges as $l\rightarrow\infty$.
Moreover,  in view of (\ref{eq::55}), (\ref{eq::57}), (\ref{eq::62}),
 (\ref{eq::61}) one has the $l$-uniform bound
\begin{equation}\label{eq::62a} 
|p_{j_0,l+1}|
<\tilde N\sum_{m=0}^l |Z_{j_0+m+2}|^{-1}
<{\tilde N(1+|n+1|^{-1}) \over j_0+1+(\pp-1)/2}.
\end{equation} 
It follows from \Eqs(\ref{eq::53}),(\ref{eq:53a})
\begin{equation}
  \label{eq::53b}
\alpha_j=
\alpha_{j_0+l}=(1+p_{j_0,l})\alpha_{j_0-1}
+
(1+p_{j_0+1,l-1} )\mu^2 Z_{j_0-\tilde\pp}^{-1}\alpha_{j_0-2}
\end{equation}
and the convergence of $\alpha$-sequence follows from the 
convergence of $p_{*,l}$ as $l\rightarrow\infty$. Then one has
\begin{equation}
  \label{eq::53c}
\lim\alpha_j-\alpha_{j_0-1}=
\lim_l p_{j_0,l} \alpha_{j_0-1}
+
(1+\lim_l p_{j_0+1,l} )\mu^2 Z_{j_0-\tilde\pp}^{-1}\alpha_{j_0-2}.
\end{equation}
The factors in front of 
 the first and second terms to the right scales as $j_0^{-1}$ and
$j_0^{-2}$, respectively.
We may therefore write down
the following inequality
\begin{equation*}
|\lim\alpha_j-\alpha_{j_0-1}|=N\max(|\alpha_{j_0-1}|,|\alpha_{j_0-2}|)j_0^{-1},
\end{equation*}
where $N$ may depend on the parameters $n,\lambda,\mu,\mykappa,\pp$
 but not on the
specific specimen of $\alpha$-sequence.
This obviously implies the inequality (\ref{eq::26}).

It has also to be noted in conclusion that the case $n=-1$
formally falling off the above speculation
does not actually correspond to an exceptional situation.
Although inequalities (\ref{eq::62})
formally fail, similar ones
differing from (\ref{eq::62})
in the values of `constant' ($j_0$-independent) terms alone
can be derived. 
The further reasoning holds true
and leads to the same conclusions.


\begin{thebibliography}{9}
 \bibitem%
 {R1}
 B.D.\ Josephson, 
 Possible new effects in superconducting tunelling,
 {Phys.\ Lett.}
  {1}(1962)
  251-253.

 \bibitem%
 {R2}
 B.D.\ Josephson,
 Coupled superconductors,
 { Rev.\ Mod.\ Phys.}
 {36} (1964) 216-220.

 \bibitem%
 {R3}
 D.E.\ McCumber,
 Effect of ac Impedance on dc Voltage-Current Characteristic
 of Superconductor Weak-Link Junctions,
 {J.\ Appl.\ Phys.}
 {39}(1968) 3113-3118.

 \bibitem%
 {R4}
 W.C.\ Stewart,
 ``Current-voltage characteristics of Josephson junctions''
 {Appl.\ Phys\ Lett.}
 {12},(1968) 277-280.


 \bibitem
 {R5}
 A.\ Barone and G.\ Paterno,
 {Physics and Applications of the Josephson Effect},
 Willey, N.Y., 1982.


 \bibitem%
 {R6}
 A.H.\ Hamilton,
 Josephson voltage standards,
 {Rev.\ of Sci.\ Instr.},
 {71} (2000) 3611-3623.

 \bibitem%
 {R7}
 S.P.\ Benz and C.A. Hamilton,
 Application of the Josephson effect to Voltage Metrology,
 {Proc IEEE}
 {92} (2004) 1617-1629.

 \bibitem%
 {R8}
 J.\ Niemeyer,
 Josephson Voltage Metrology,
 in: {Abstract Booklet EUCAS 2005}, Vienna (2005) 83-84. 

 \bibitem%
 {R9}
 A. Kemooinen, J. Nissil\"a, K. Ojasalo, J. Hassel, A. Manninen,
 P. Helist\"o
 and H. Sepp\"a,
 AC Voltage Standard based on an Externally-Shunted SIS Josephson
 Junction Array,
 in: Abstract Booklet EUCAS 2005, Vienna (2005) 336. 

 \bibitem%
 {R10}
 R.L. Kautz,
 Noise, chaos, and Josephson voltage standard,
 Rep. Prog. Phys.,
 {59} (1996) 935-992.

 \bibitem
 {R11}
 S.I.\  Tertychniy,
 ``Long-term behavior 
 of solutions to the
 equation $\dot \phi + \sin\phi=f$ 
 with periodic $f$ 
 and the modeling of dynamics 
 of
 overdamped Josephson junctions''
 Preprint
 {Arxiv:math-ph/0512058} (2005).
 
\bibitem
 {R11a}
 S.I.\  Tertychniy,
``The modeling of a Josephson junction and Heun polynomials''
 Preprint
 {Arxiv:math-ph/0601064} (2006).

 \bibitem
 {R12}
 V.M.\ Buchstaber, O.V.\ Karpov  and S.I.\ Tertychniy,
 Quantum Josephson D/A converter driven by trains of short $2\pi$-pulses,
 in: {Conference Digest CPEM 2002}, Ottawa, (2002) 502-503
 (2002).

 \bibitem%
 {R13}
 D.\ Schmidt, G. Wolf,
 Double confluent Heun equation, in:
 Heun's differential equations,
 Ronveaux (Ed.)
 Oxford Univ. Press, Oxford, N.Y.,  (1995).\, Part C.

 \bibitem
 {R14}
 S.Yu.\ Slavyanov and W.\ Lay,
 Special functions: A Unified Theory Based on Singularities,
 Nevskiy dialect, SPb, (2002), in Russian; 
 English edition: 
 Oxford Univ. Press, Oxford, N.Y., (2000).


















\end{thebibliography}
\end{document}